\documentclass[superscriptaddress]{revtex4}

\usepackage{graphicx}
\def\TeV{\ifmmode {\,\mathrm{ Te\kern -0.1em V}}\else
                   \textrm{Te\kern -0.1em V}\fi}%
\def\GeV{\ifmmode {\,\mathrm{ Ge\kern -0.1em V}}\else
                   \textrm{Ge\kern -0.1em V}\fi}%
\def\MeV{\ifmmode {\,\mathrm{ Me\kern -0.1em V}}\else
                   \textrm{Me\kern -0.1em V}\fi}%
\def\keV{\ifmmode {\,\mathrm{ ke\kern -0.1em V}}\else
                   \textrm{ke\kern -0.1em V}\fi}%
\def\eV{\ifmmode  {\,\mathrm{ e\kern -0.1em V}}\else
                   \textrm{e\kern -0.1em V}\fi}%
\let\gev=\GeV

\newcommand{\WW}    {\mathrm{W}^+\mathrm{W}^-}
\newcommand{\ALR}{\mbox{$A_{\rm {LR}}$}}

\newcommand{\ee}    {\mathrm{e}^+\mathrm{e}^-}

\newcommand{\ppl}  {{\cal P}_{e^+}}
\newcommand{\pmi}  {{\cal P}_{e^-}}
\newcommand{\ppm}  {{\cal P}_{e^\pm}}

\newcommand{\MZ}      {m_{\mathrm{Z}}}
\newcommand{\MW}      {m_{\mathrm{W}}}

\newcommand{\GZ}      {\Gamma_{\mathrm{Z}}}
\newcommand{\GW}      {\Gamma_{\mathrm{W}}}

\newcommand {\cAe} {\mbox{$\cal A_{\rm e}$}}
\newcommand {\cAb} {\mbox{$\cal A_{\rm b}$}}

\newcommand {\cAq} {\mbox{$\cal A_{\rm q}$}}
\newcommand {\Rb}   {R_{\mathrm{b}}}
\newcommand {\so}   {\sigma_0^{\rm{had}}}
\newcommand {\stl}  {\sin^2 \theta_{{\rm eff}}^\ell}
\newcommand{\Ghad}       {\Gamma_{\mathrm{had}}}
\let \sweff=\stl
\newcommand{\de}{\delta}
\newcommand{\De}{\Delta}
\newcommand{\al}{\alpha}
\newcommand{\BE}{\begin{equation}}
\newcommand{\EE}{\end{equation}}
\newcommand{\BC}{\begin{center}}
\newcommand{\EC}{\end{center}}
\newcommand{\Stope}{\tilde{t}_1}

\newcommand{\tst}{\theta_{\tilde{t}}}
\newcommand{\mste}{m_{\tilde{t}_1}}
\newcommand{\mstz}{m_{\tilde{t}_2}}
\begin{document}
\bibliographystyle{revtex}


\title{Positron polarisation and low energy running at a Linear Collider}



\author{J. Erler}
\email[]{erler@ginger.hep.upenn.edu}
\affiliation{Department of Physics and Astronomy, University of Pennsylvania,
             Philadelphia, PA 19146-6396, USA}

\author{K. Fl\"ottmann}
\email[]{Klaus.Floettmann@desy.de}
\affiliation{DESY, Hamburg \& Zeuthen, Germany}

\author{S. Heinemeyer}
\email[]{Sven.Heinemeyer@bnl.gov}
\affiliation{HET Physics Dept., Brookhaven Natl.\ Lab., NY, USA}

\author{K. M\"onig}
\email[]{Klaus.Moenig@desy.de}
\affiliation{DESY, Hamburg \& Zeuthen, Germany}

\author{G. Moortgat-Pick}
\email[]{gudrid@mail.desy.de}
\affiliation{DESY, Hamburg \& Zeuthen, Germany}

\author{P. C. Rowson}
\email[]{rowson@slac.stanford.edu}
\affiliation{Stanford Linear Accelerator Center, USA}

\author{E. Torrence}
\email[]{torrence@physics.uoregon.edu}
\affiliation{Department of Physics, University of Oregon, USA}

\author{G. Weiglein}
\email[]{Georg.Weiglein@durham.ac.uk}
\affiliation{Institute for Particle Physics Phenomenology, Durham, UK}

\author{G. W. Wilson}
\email[]{graham@fnal.gov}
\affiliation{Department of Physics and Astronomy, University of Kansas, Lawrence, KS 66045, USA}


\date{\today}

\begin{abstract}
%
The physics potential of an $\ee$ linear collider can be significantly
enhanced if both the electron and positron beams are polarised.
Low energy running at the Z-resonance or close to the W-pair threshold 
is particularly attractive with polarised positrons.
This note discusses the experimental aspects and physics opportunities 
of both low energy running and positron polarisation.
\end{abstract}

\maketitle

\section{Introduction}
\label{sec:intro}
An $\ee$ linear collider offers many possible options to enhance
the baseline program \cite{ref:tdr_phys,ref:orange}. 
Already in its basic running mode the electron beam
will be polarised to around 80\% with a strained photocathode
technology similar to that used at the SLC.
As two additional options it should also be possible to polarise
the positrons and to run at lower energies around the Z-pole and
the W-pair threshold.

Positron polarisation enables some genuinely new  measurements,
especially in Supersymmetry \cite{gudi}. In addition, polarised positrons
improve the measurement of the beam polarisation due to favourable error
propagation and due to the possibility of measuring the polarisation 
directly using $\ee\to\WW$ events \cite{pol_note}.

Electroweak tests, already performed at LEP and SLC 
can be repeated with much higher precision with the linear collider
running at low energy~\cite{kmogigaz,gigaz}. The largest progress can be
achieved in the measurement of the effective weak mixing angle using
the left right asymmetry. This measurement will be completely limited by 
polarisation systematics unless positron polarisation is available.
In the measurement of the W-mass the background can be controlled by measuring
the left-right asymmetry near the W-pair threshold. Also this requires 
a fairly accurate polarisation measurement that can be helped by positron
polarisation.

\section{Positron polarisation}
Simultaneous electron and positron beam polarisation results in six 
principal advantages: 
(1) higher effective polarisation,  
(2) suppression of background,
(3) enhancement of event rates,
(4) increased sensitivity to non-standard couplings, 
(5) fixing quantum numbers of new particles and 
(6) improved accuracy in measuring the polarisation \cite{gudi,peter}. 

The fact that 
highly polarised electron beams are achievable in a linear collider has 
already been demonstrated at the SLC, and there is every reason to expect
that electron polarisations $\pmi$ 
in excess of $80\%$ will be possible at future 
linear colliders. Furthermore, methods for achieving 40--60\% positron 
polarisation $\ppl$ have been proposed and are currently under development.

\subsection{Technical issues concerning positron polarisation}
Compared to polarised electron sources, the technical hurdles for
positron polarisation are significant. A fundamental difference
is that the production of each positron requires 10-100 MeV photons,
rather than the few eV photons per electron at an electron photocathode.
In addition, the yields are typically an order of magnitude worse for
positrons than for photoproduced electrons.   Nevertheless, three different
technical approaches for polarised positron production have been
discussed in the literature: 1) bremsstrahlung pair production
with a polarised electron beam, 2) Compton
backscattering of photons from a high energy polarised electron beam,
with subsequent photo-production of positrons, and 3) polarised photon
production using a high energy electron beam in a helical undulator, with
subsequent photo-production of positrons.

The first method, where for example a 50 MeV electron beam is
incident on a 0.1 radiation length target, and where the produced
positrons with energy larger than 25~MeV are captured and transported,
would in principle produce a positron polarisation of 50\% \cite{Potylitsin}.
But the efficiency is low and a beam power of order 1.5 MW would be required.
As this power is comparable to the expensive (of order 1\,G\$)
CEBAF beam, the first method has been deemed impractical.
The second technique is attractive because the positron polarisation
would be controllable pulse-to-pulse by changing the circular polarisation
of the laser.  However, a dedicated high current 6 GeV electron
linac and a complex laser system consuming a tremendous amount of laser
power would
be needed - in one design, a system of about 50 ${\rm CO}_2$ lasers
using a ``wall plug'' power of about 20 MW, in order to achieve 50-60\%
positron polarisation \cite{Omori}. The third method would use $\sim 200$ 
meters of helical undulator magnet
through which the full-energy electron beam is passed, producing polarised
photons.
A collimated fraction of the photon beam ($\sim20\%$)
is directed onto a target, and positrons of energy larger than 
$15 \MeV$ are retained producing 
a 60\% polarised beam \cite{Balakin,Flottman}.
A low emittance electron beam is required, and hence the 
post-collision beam probably cannot be used.
TESLA proposes a similar design also for the non-polarised
positron source \cite{tdr_machine}.  For high energy running the same electron
beam is used for positron creation and for physics. This scheme reduces the
energy of the colliding ${\rm e}^-$ beam by ${\cal O}(1\%)$ and increases the
energy spread from 0.05\% to 1.5\% which is considered acceptable.
Alternately, separate electron bunches could be used to produce
positrons preserving the energy spread of the colliding beam, 
but at the cost of ultimate luminosity.
For GigaZ running one part of the electron arm is used to produce the
$\sim 50 \GeV$ physics beam while the other part accelerates a $\sim 200 \GeV$
beam to produce the positrons.
Technical issues arise regarding the construction of this first-of-a-kind
undulator and the associated photon collimation and positron capture 
systems, and the cost of such a positron source will certainly be high.

\subsection{Physics benefits of positron polarisation at high energies}
In the limit of vanishing electron mass,
SM processes in the s--channel are initiated by
electrons and positrons polarised in the same direction, i.e. $e^+_Le^-_R$ (LR)
or $e^+_Re^-_L$ (RL), where the first (second) entries 
denote helicities of the corresponding particles. 
This result follows from the vector nature of
$\gamma$ or $Z$ couplings (helicity--conservation).
In the following the convention will be used that, 
if the sign is explicitly given, $+$ $(-)$ polarisation
corresponds to R (L) chirality with helicity $\lambda=+\frac{1}{2}$ 
($\lambda=-\frac{1}{2}$) for both electrons and positrons.
For these processes positron polarisation provides no fundamentally new 
information. However, choosing the suitable beam polarisation can 
significantly enhance rates and suppress background.
In theories beyond the SM 
both
(LL) and (RR) configurations for s-channel contributions are also allowed and
so the polarisation of both beams offers a powerful tool, in addition to 
enhancing rates and suppressing SM backgrounds, 
for analysing the coupling structure of the underlying theory. 

A short overview of the polarisation effects of a future
linear collider is given in \cite{gudi}.
Since, however, one of the main advantages of having 
positron polarisation is related to the study of SUSY particles two 
examples for SUSY processes will be discussed here in more detail.

\noindent{\bf Higgs physics}\\
Higgs production at a LC occurs mainly via $WW$ fusion, 
$e^+e^-\to H \nu \bar{\nu}$, and Higgsstrahlung $e^+ e^-\to HZ$. 
For a light Higgs of about
$m_H\le 130$~GeV both processes have comparable 
cross sections at a LC with $\sqrt{s}=500$~GeV.
Beam polarisation can help to measure the $HZZ$ and the $HWW$ coupling 
separately, e.g.~via suppression of the $WW$ background 
(and the signal of $WW$ fusion) and enhancement of the $HZ$ contribution with 
right polarised electrons and left polarised positrons. 
Furthermore, beam polarisation reduces considerably the error when 
determining the Higgs couplings.

\noindent{\bf Electroweak physics}\\
At a LC it is possible to test the SM and its prediction
for couplings and mixing angles with unprecedented accuracy.
a) high $\sqrt{s}$: In order to test the SM with high precision one can 
carefully study triple gauge boson couplings in the process 
$e^+ e^-\to W^+ W^-$ by measuring the angular
distribution and polarisation of the $W^{\pm}$'s. 
Simultaneously fitting all of the couplings using unpolarised cross
sections results in a strong 
correlation between the $\gamma-$ and $Z-$couplings whereas 
polarised beams are well suited to separate these couplings.
However, the statistical error in the gauge 
couplings is small compared to the error due to 
the experimental uncertainty of $\pmi$.
Simultaneously polarised $e^+$ and $e^-$ beams reduce this error
significantly in the polarisation measurement.\\
b) At GigaZ, $\sqrt{s}=m_Z$, 
an order--of--magnitude improvement in the accuracy of
the determination of $\stl$ may well be 
possible when using the Blondel Scheme, as discussed in 
section \ref{sec:stlgigaz}.

\noindent{\bf QCD physics}\\
a) The LC offers the possibility of testing QCD at high energy scales 
with very high accuracy. Besides the improvement of rates and 
suppression of e.g.~$WW$ background in QCD in general the simultaneous 
polarisation of both beams leads 
in particular to a precise determination of top properties
as well as to extreme limits of top flavour changing neutral (FCN) 
couplings. \\
b) The LC can also collide electrons with electrons and
the possibility of
$\gamma \gamma$ and $e^- \gamma$ collisions is under active study.
All three ``novel'' collision modes could be used to
study polarised
structure functions (PSF) of photons. 
However, even the standard $e^+ e^- $ mode 
can be used to
gain information on
PSF if one uses highly polarised $e^+$ and 
$e^-$ beams in the process 
$e^+ e^-\to \gamma \gamma + e^+ e^-\to \mbox{Di-jets}+e^+e^-$.
Since effects of depolarisation tend to be 
large at the $e \gamma$ vertex one needs highly polarised $e^-$ and $e^+$ 
beams to get first experimental hints on polarised PSF.

\noindent{\bf Alternative Theories}\\
a) Using simultaneous polarisation of both beams increases the 
effective polarisation from e.g.~$80\%$ to $95\%$ (using the 
configuration $(80\%,60\%)$) and leads to a higher effective luminosity. 
Beam polarisation is therefore 
a helpful tool to enlarge the discovery reach of $Z'$, 
$W'$ and to discriminate between different contact interactions. \\
b) In the direct search for extra dimensions,
$e^+ e^-\to \gamma G$, beam polarisation enlarges the discovery reach for 
the scale $M_D$, and is in particular crucial for enhancing the signal (S) and
suppressing the dominant background  (B) $e^+ e^-\to \nu \bar{\nu} \gamma$.
The main background is dominated by left--handed couplings and consequently the ratio 
$\frac{S}{\sqrt{B}}$ increases by a factor 2.1  
when using $(\pmi,\ppl) = (+0.8,0)$ and by a factor 4.4 when using 
$(+0.8,-0.6)$ for the study described in \cite{GWW_ED}.

\noindent {\bf Supersymmetry}\\
a) Simultaneous polarisation of both beams is absolutely needed for 
establishing the partnership between electron states and selectron states in
Supersymmetry (SUSY), where the scalar particles get associated  
chiral quantum numbers of their SM partners:
$e^-_{L,R} \leftrightarrow \tilde{e}^-_{L,R}$,  
$e^+_{L,R} \leftrightarrow \tilde{e}^+_{R,L}$.

\begin{figure}[htb]
\setlength{\unitlength}{1mm}
\begin{center}
\begin{picture}(83,73)
\put(-.2,0){\includegraphics{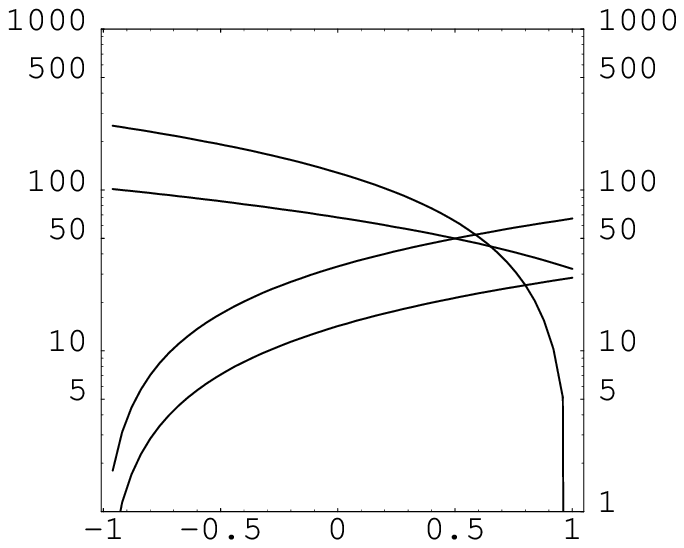}}
\put(-1.2,0){\includegraphics{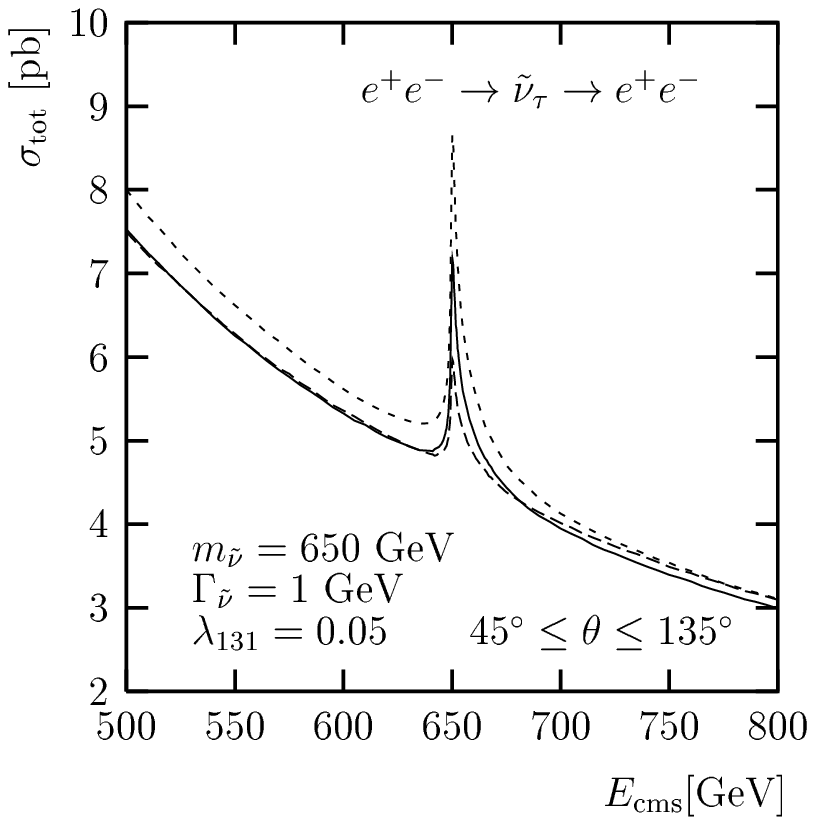}}
\put(78,63){b)}
\put(-25,60){\makebox(0,0)[bl]{{
   a) $\sigma(e^- e^+\to \tilde{e}^-_{L,R} \tilde{e}^+_{L,R}$)/fb}}}
\put(-25,49){$\tilde{e}^{-}_{L} \tilde{e}^{+}_{R}$}
\put(13,40){$\tilde{e}^{-}_{L} \tilde{e}^+_{L}$}
\put(0,23){$\tilde{e}^-_{R} \tilde{e}^+_{L}$}
\put(-28,36){$\tilde{e}^-_{R} \tilde{e}^+_{R}$}
\put(20,-3){$P_{e^+}$}
\end{picture}
\end{center}
\caption[]{a) Production cross sections as a function of $P_{e^+}$ for
    $\sqrt{s} = 400$~GeV, $P_{e^-}=-0.8$, $m_{\tilde{e}_R}=137.7$ GeV,
  $m_{\tilde{e}_L}=179.3$ GeV, $M_2=156$~GeV, $\mu=316$~GeV and
  $\tan\beta=3$, $\mu=316$~GeV and $\tan\beta=3$.
  ISR corrections and beamstrahlung are included \cite{slep_01}.
b) Sneutrino production in R--parity violating model:
Resonance production of
$e^+ e^-\to \tilde{\nu}$ interfering with Bhabha scattering 
for different configurations of 
beam polarisation: unpolarised case (solid), 
$P_{e^-}=-80\%$ and $P_{e^+}=+60\%$ (long-dashed), 
$P_{e^-}=-80\%$ and $P_{e^+}=-60\%$ (short-dashed) \cite{Spiesi} 
          }
\label{fig_slep}
\end{figure}

The s--channel only allows incoming particle/antiparticle
pairs $e_L^- e_R^+$ and $e_R^- e_L^+$ due to helicity conservation, whereas in
the t--channel all possible beam configurations are allowed:
\begin{eqnarray}
\mbox{s-- and t--channel}&:&
e_L^- e_R^+ \to \tilde{e}^-_L \tilde{e}^+_L, \tilde{e}^-_R 
\tilde{e}^+_R, \quad\mbox{and}\quad
e_R^- e_L^+ \to \tilde{e}^-_L \tilde{e}^+_L, \tilde{e}^-_R 
\tilde{e}^+_R ~, \label{eq_2b}\\
\mbox{t--channel}&:&
e_L^- e_L^+ \to \tilde{e}^-_L \tilde{e}^+_R, \quad\phantom{and}\quad
\mbox{and}\quad
e_R^- e_R^+ \to \tilde{e}^-_R \tilde{e}^+_L ~.
\end{eqnarray}  
Polarised cross sections including ISR and beamstrahlung
for the different selectron pairs at $\sqrt{s}=400$~GeV close to the 
production threshold are shown in Fig.~\ref{fig_slep} a). When 
using $\pmi=-80\%$ and variable $\ppl$, one sees
that even for $\ppl=-40\%$~(LL) the highest rates are those for the pair 
$\tilde{e}^-_L\tilde{e}^+_R$. Its rate is more than a factor two 
larger than for all other pairs. The two particles can now 
be separated via charge identification. For these analyses the simultaneous
polarisation of the positron beam is absolutely needed.
It should be noted that this test of properties is not possible
when running at energies far above the corresponding production
threshold due to the event kinematics\cite{slep_01}.
b) In SUSY models all coupling structures consistent with Lorentz invariance
should be considered. Therefore it is also possible to get appreciable 
event rates for polarisation configurations that are unfavourable for SM 
processes. Therefore one example in R--parity violating SUSY,
$e^+ e^-\to \tilde{\nu} \to e^+ e^-$, is studied, which is 
characterised by the exchange of a scalar particle in the direct channel.
The main background to this process is Bhabha scattering. 
A study \cite{Spiesi} was made for $m_{\tilde{\nu}}=650$~GeV,
$\Gamma_{\tilde{\nu}}=1$~GeV, with an angle cut of 
$45^{0}\le \Theta\le 135^{0}$ and a lepton--number violating coupling 
$\lambda_{131}=0.05$ in the R--parity violating Lagrangian
${\cal L}_{\not R}\sim \sum_{i,j,k}\lambda_{ijk} L_i L_j E_k$. 
Here $L_i$ denotes the lepton doublet super fields while $E_k$
corresponds to the lepton singlets.
In Fig.~\ref{fig_slep} b) the resonance 
curve for the process, including the complete SM--background is given. 
The cross section $\sigma(e^+ e^-\to e^+ e^-)$ including
$\sigma( e^+ e^-\to \tilde{\nu}\to e^+ e^-)$ gives i) 7.17 pb
(including Bhabha--background: 4.50 pb) for the unpolarised case, ii) 7.32 pb 
(including Bhabha--background: 4.63 pb) for $P_{e^-}=-80\%$ and
iii) 8.66 pb (including Bhabha--background: 4.69 pb) for
$P_{e^-}=-80\%$, $P_{e^+}=-60\%$. This means that only the 
electron polarisation  
enhances the signal only slightly by about 2\%, whereas the simultaneous
polarisation of both beams with $(-80,-60)$ 
produces a further increase by about 20\%. 
This configuration of beam polarisations, which strongly suppresses pure SM 
processes, allows one to perform fast diagnostics 
for this R--parity violating process.\\
c) Simultaneously polarised $e^-$ and $e^+$ beams lead to 
the additional enhancements of rates for specific processes with
a corresponding suitable choice of beam polarisation. Such an enhancement 
can not be reached if only electron beams are polarised even if 100\% 
polarisation were provided and it can be decisive for the discovery of 
SUSY particles e.g.~in extended SUSY models when particularly
small rates have to be taken into account. 
 
Furthermore beam polarisation plays a decisive role in the discovery
of new particles (by enhancing their production cross sections) and 
in particular in the determination of SUSY parameters.
The analysis of polarised cross sections for the process
$e^+ e^- \to \tilde{t}_1 \bar{\tilde{t}_1}$ leads to a precise
determination of the stop mixing angle, see
Sect.~\ref{subsec:testsusy}. 
The determination of the fundamental SUSY parameters (even if CP--violating) 
can be derived from the chargino and neutralino sector
when measuring masses and polarised cross sections (see e.g.\
Ref.~\cite{gudi_jan} and references therein.)
In this context the simultaneous polarisation of both beams will lead 
to a higher accuracy.
Moreover after the determination of the 
MSSM parameters it will be possible to test experimentally at a LC
the fundamental SUSY prediction that the Yukawa couplings, 
$g_{_{\tilde{W}}}$ and $g_{_{\!\tilde{B}}}$, are identical to the SU(2) and 
U(1) gauge couplings $g$ and $g'$ very accurately.
 
Determining the model parameters with high accuracy and exploiting numerous 
consistency relations which are based on analytical calculations provide a 
powerful tool to illuminate the underlying structure of the supersymmetric 
model.\\

\noindent 
To summarize,
the clean and fundamental nature of $e^+ e^-$ collisions in a linear 
collider is 
ideally suited for the search for new physics, and the determination of both 
Standard Model and New Physics couplings with high precision.
Polarisation effects will play a crucial role in these processes. 
We have presented numerous examples 
that simultaneous polarisation of both beams can 
significantly expand the accessible physics opportunities for 
a complete reconstruction of the underlying theory with high accuracy.

\section{Low energy running}

In principle all measurements done at LEP and SLC can be repeated at
the linear collider, however with much smaller statistical errors.
In around 100 running days $10^9$ Z-decays can be collected, about 50 times
the LEP or 2000 times the SLC statistics. 
Two areas are of special interest for the linear collider: 
electroweak and B-physics. For all measurements related to Z-couplings there
exists no real alternative to Z-pole running and the W-pair threshold region
seems to be the best place for a precise W-mass measurement. 
The situation is more complex for B-physics. In $10^9$ hadronic Z decays there
are about $4 \cdot 10^8$ b-hadrons. This data sample is comparable to the $\ee$
B-factories running at the $\Upsilon(4S)$ and much smaller than the samples
expected at the TEVATRON or the LHC. 
However, contrary to the $\Upsilon(4S)$ all
b-hadron species are produced and contrary to the hadron machines all events
can be reconstructed. In addition, the b-quarks are highly polarised at 
production and, due to the large forward-backward asymmetry with polarised 
beams, the production-charge tagging can be done with good purity from the
b-direction only. A more detailed discussion of B-physics with Z-running
at a linear collider can be found in \cite{e2report}.

\subsection{Machine issues}
The present linear collider designs can deliver luminosities of
${\cal L} \sim 5 \cdot 10^{33} {\rm cm}^{-2} {\rm s}^{-1}$ at 
$\sqrt{s}\sim \MZ$
and
${\cal L} \sim 10^{34} {\rm cm}^{-2} {\rm s}^{-1}$ at 
$\sqrt{s}\sim 2 \MW$.

 The energy
loss due to beamstrahlung for colliding particles is around
$0.05\%-0.1\%$ at $\MZ$ and
about a factor of four higher at
$2 \MW$.
The depolarisation in the interaction region is
basically negligible. Sacrificing some luminosity the beamstrahlung can
be reduced substantially, for example by a factor three for a
luminosity loss of a factor two.

Apart from the beamstrahlung there are several other effects that
influence the precision of the measurements: 
\begin{itemize}
\item The mean energies of the two
  beams have to be measured very precisely. 
  A precision relative to the Z mass of around $10^{-5}$ might be needed.
\item The beam energy spread of the machine plays a crucial role in
  the measurement of the total width of the Z. If the shape of the
  distribution is known it can be measured from the acolinearity of
  Bhabha events in the forward region as long as the energies of the
  two colliding particles are not strongly correlated.
\item With the high luminosities planned, the Z-multiplicity in a train
  becomes high. This can influence Z-flavour tagging or even Z counting.
\end{itemize}
The two main designs, X-band and superconducting, differ in some
aspects relevant for running at energies around the
Z-pole~\cite{ref:tdr_phys,ref:orange}. 
For the X-band design a bunch train
contains 190 bunches with 1.4\,ns bunch spacing. In this case more
than half of the Z-bosons are produced in the same train with at least
one other Z, and the ability to separate events must be studied in detail.
A TESLA train contains 2820 bunches with 337\,ns bunch spacing. In
this case event overlap is not a problem, but the requirements for
the data acquisition system are higher. 
The smaller wakefields in the superconducting machine should
reduce the beam energy spread, and the larger bunch spacing should
result in a smaller energy difference between the bunches in a
train. 
If the beam spectrometer has some ability to identify individual
bunches, however, this latter effect is not a serious problem
for either design.

In order not to inhibit the electroweak precision measurements already
in the LC design it has to be assured that suitable space in the
beam delivery system for very precise beam energy measurement and
polarimetry is provided or that the beam energy measurement is directly
incorporated into the final focus magnet system. 
A measurement of these quantities behind the IP will also be very helpful.

\subsection{Electroweak observables}
There are four classes of electroweak observables that can be
measured at a linear collider during low energy running at the
Z-resonance or W-pair threshold:
\begin{itemize}
\item observables related to the partial widths of the Z, measured in
  a Z-resonance scan;
\item observables sensitive to the effective weak mixing angle;
\item observables using quark flavour tagging;
\item the W boson mass and width.
\end{itemize}
Table \ref{tab:line} summarises the present precision and the
expectations for the linear collider for these quantities.
\begin{table}
\begin{center}
\begin{tabular}[c]{|c|c|c|}
\hline
 & LEP/SLC/Tev \cite{ref:lepew} & GigaZ \\
\hline
$\stl$ & $0.23146 \pm 0.00017$ & $\pm 0.000013$\\
\hline
\multicolumn{3}{|l|}{Z lineshape observables:}\\
\hline
$\MZ$ & {$ 91.1875 \pm 0.0021 \GeV$} & {$ \pm 0.0021 \GeV$} \\
$\alpha_s(\MZ^2)$ & {$ 0.1183 \pm 0.0027 $} & {$ \pm 0.0009$} \\
$\Delta \rho_\ell$ & {$ (0.55 \pm 0.10 ) \cdot 10^{-2}$} 
& {$ \pm 0.05\cdot 10^{-2}$}  \\
$N_\nu$ & {$ 2.984 \pm 0.008 $} & {$ \pm 0.004 $} \\
\hline
\multicolumn{3}{|l|}{heavy flavours:}\\
\hline
$\cAb$ & $0.898 \pm 0.015$   &$\pm 0.001$ \\
$\Rb^0$ &$0.21653 \pm 0.00069$ & $\pm 0.00014$  \\
\hline
\multicolumn{3}{|l|}{W boson:}\\
\hline
$\MW$ & $80.451 \pm 0.033 \GeV$ & $\pm 0.007$ \\
$\GW$ &  $2.114 \pm 0.076 \GeV$ & $\pm 0.004$ \\
\hline
\end{tabular}
\end{center}
\caption{
Possible improvement in the electroweak physics quantities 
at a linear collider.
The W-boson mass will likely be known to better than 15 MeV after
the LHC begins.
For $\alpha_s$ and $\Delta \rho_\ell$, $N_\nu=3$ is assumed.
}
\label{tab:line}
\end{table}

\subsubsection{Observables from the Z-line scan}
{}From a scan of the Z-resonance curve the following quantities are measured:
\begin{itemize}
\item the mass of the Z ($\MZ$);
\item the total width of the Z ($\GZ$);
\item the hadronic pole cross section 
  ($\so = \frac{12 \pi}{\MZ^2} \frac{\Gamma_e \Ghad}{\Gamma_Z^2} $);
\item the ratio of the hadronic to the leptonic width of the Z
($R_\ell = \frac{\Ghad}{\Gamma_l} $). 
\end{itemize}

{}From these parameters two
interesting physics quantities, the radiative correction parameter
normalising the Z leptonic width, $\Delta \rho_\ell$, and the strong
coupling constant, $\alpha_s$, can be derived.

All observables are already at LEP systematics limited, so that the
improvement in the statistical errors is not an issue.
{}From LEP, $\MZ$ is known to $2\cdot 10^{-5}$, while the other three
parameters are all known to about $10^{-3}$. To improve on $\alpha_s$ and
especially on $\Delta \rho_\ell$, it would be best to improve all
three parameters.  This implies that one has to scan
for $\GZ$, needs the hadronic and leptonic
selection efficiencies for $R_\ell$, and in addition the absolute
luminosity for $\so$.
Due to the better detectors and the higher statistics that can be used
for cross checks, the errors on the selection efficiency and the
experimental error on the luminosity might be improved by a factor three
relative to the best LEP experiment \cite{ref:marc} 
while it is not clear whether the theory error on the
luminosity can be improved beyond its present value of $0.05\%$.
These errors would improve the precision on $R_\ell$ by a factor of
four and $\so$ by 30\%.

For an improvement in $\GZ$ a very precise point to point measurement
of the beam energy is needed while the absolute calibration can be 
obtained from the $\MZ$ measurement at LEP.
With a M{\o}ller spectrometer a precision of $10^{-5}$ of the beam
energy relative to $\MZ$ is realistic, leading to a potential
improvement of a a factor two in $\GZ$. However, because the second
derivative of a Breit-Wigner curve at the maximum is rather large,
$\GZ$ and $\so$ are significantly modified due to beamstrahlung and
beam energy spread. For the TESLA parameters 
the fitted $\GZ$ is increased by about $60 \MeV$ and $\so$ is
decreased by 1.8\%, where the majority comes from the beamspread.
The beamspread thus needs to be understood to about 2\% in order not
to limit the precision of $\Delta \rho_\ell$.
There is the
potential to achieve this precision with the acolinearity measurement
of Bhabha events \cite{ref:bsmeas} or to use at least five scan
points and fit to the beamspread, but both options need further studies.

\subsubsection{The effective weak mixing angle}
\label{sec:stlgigaz}
If polarised beams are available, the quantity that is by far most sensitive to
the weak mixing angle is the left-right asymmetry:
\begin{eqnarray}
\label{eq:alrdef}
\ALR & = & \frac{1}{{\cal P}}\frac{\sigma_L-\sigma_R}{\sigma_L+\sigma_R}\\
     & = &   \cAe \nonumber \\
     & = & \frac{2 v_e a_e}{v_e^2 +a_e^2} ~,\nonumber \\
  {v_e}/{a_e} & = & 1 - 4 \stl ~,\nonumber
\end{eqnarray}
independent of the final state.

Details of this measurement are reported in \cite{kmogigaz,peter_mike}.
With $10^9$ Zs, an electron polarisation of 80\% and no positron
polarisation the statistical error is $\Delta \ALR = 4 \cdot 10^{-5}$.
The error from the polarisation measurement is
$\Delta \ALR/\ALR = \Delta {\cal P}/{\cal P}$.
At SLC $\Delta {\cal P}/{\cal P}=0.5\%$ has been reached \cite{alrsld}. 
With some optimism
a factor two improvement is possible leading to 
$\Delta \ALR = 3.8 \cdot 10^{-4}$. This is already more than a factor
five improvement relative to the final SLD result and almost a factor
four compared to
the combined LEP/SLD average on $\stl$.

\begin{figure}[htbp]
  \begin{center}
    \includegraphics[height=6cm,bb=0 6 567 472]{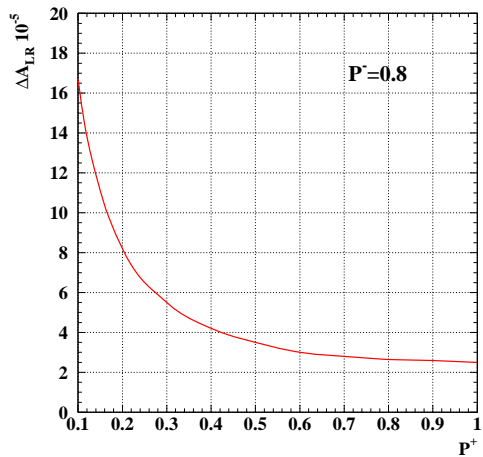}
  \end{center}
  \caption{Error of $\ALR$ as a function of the positron
    polarisation for a luminosity corresponding to $10^9$ unpolarised Zs.}
  \label{fig:polerr}
\end{figure}

If positron polarisation is available there is the potential to go
much further using the Blondel scheme \cite{ref:alain}. 
The total cross section with both beams being polarised is given as
  $\sigma \, = \, \sigma_{\rm unpol} \left[ 1 - \ppl \pmi + \ALR (\ppl - \pmi) \right]$.
If all four helicity combinations are measured $\ALR$ can be determined
without polarisation measurement as
\[
\ALR \, = \, \sqrt{\frac{
    ( \sigma_{++}+\sigma_{-+}-\sigma_{+-}-\sigma_{--})
    (-\sigma_{++}+\sigma_{-+}-\sigma_{+-}+\sigma_{--})}{
    ( \sigma_{++}+\sigma_{-+}+\sigma_{+-}+\sigma_{--})
    (-\sigma_{++}+\sigma_{-+}+\sigma_{+-}-\sigma_{--})}} ~.
\]
Figure \ref{fig:polerr} shows the error of $\ALR$ as a function on the
positron polarisation.  For $\ppl > 50\%$ the dependence is relatively weak.
For $10^9$ Zs a positron polarisation of 20\% is better than a
polarisation measurement of $0.1\%$ and electron polarisation only.

However, polarimeters for relative measurements are still needed. The
crucial point is the difference between the absolute values of the
left- and the right-handed states.
If the two helicity states for electrons and positrons are written as 
$\ppm = \pm |\ppm| + \delta \ppm$ the
dependence is $\rm{d} \ALR / \rm{d} \delta \ppm \approx 0.5$.
One therefore needs to understand $\delta \ppm $ to $< 10^{-4}$.
If polarimeters with at least two channels
with different analysing power are available, not only the analysing
powers but also some internal asymmetries in the polarimeters can be
fitted to the data.

Due to $\gamma-Z$ interference, the dependence of $\ALR$ on the beam
energy is given by $\rm{d} \ALR / \rm{d} \sqrt{s} = 2 \cdot 10^{-2}/\GeV$. The
difference $\sqrt{s} - \MZ$ thus needs to be known to $\sim 10 \MeV$
to match the measurement with electron polarisation only, and to $\sim 1\MeV$
if polarised positrons are available, the same precision as for the foreseen
$\GZ$ improvement.
For the same reason, beamstrahlung shifts $\ALR$ by $\sim 9 \cdot 10^{-4}$
(TESLA design), so its uncertainty should only be a few percent.
If beamstrahlung is identical in the Z-scan that calibrates the
beam energy it gets absorbed in the energy calibration, so that
practically no corrections are needed for $\ALR$.
How far the beam parameters can be kept constant during the scan and
how well the beamstrahlung can be measured still needs further studies.
However, for $\ALR$ only the beamstrahlung and not the beamspread matters.
If the beamstrahlung cannot be understood to the required level in the
normal running mode one can still go to a mode with lower beamstrahlung
increasing the statistical error or the running time.

For the interpretation of the data it will be assumed that 
$\Delta \ALR =10^{-4}$ is possible, 
corresponding to $\Delta \stl = 0.000013$. 
However, it has to be kept in mind that this error will increase by a
factor four if no positron polarisation is available.

\subsubsection{Observables with tagged quarks}
Using quark tagging in addition to the observables already discussed,
the partial widths and forward-backward asymmetries for b- and
c-quarks can also be measured. 
These observables are sensitive to vertex corrections at the
${\rm Zq\bar{q}}$ vertex and to new Born-level effects that alter the SM 
relations between quarks and leptons. Especially the ${\rm Zb\bar{b}}$
vertex is very
interesting, since the b-quark is the isospin-partner of the heavy top
quark that plays a special role in many extensions of the SM.
It should be noted that the leading vertex corrections are enhanced by
the square of the top quark mass.

Up to now only estimates for b-quark observables 
exist \cite{kmogigaz,Ali:2000cq}.
For the ratio of the Z partial widths to b-quarks and to hadrons $(\Rb)$
an improvement of a factor five to the LEP/SLD average is
possible. This improvement is due to the much better b-tagging compared to
LEP, which allows for a higher purity and a smaller energy-dependence
reducing the hemisphere correlations.

The forward-backward asymmetry with unpolarised beams measures the
product of the coupling parameters for the initial state electrons and
the final state quarks $A_{\rm FB}^q = \frac{3}{4}\cAe \cAq$ while the left-right
forward-backward asymmetry with polarised beams is sensitive to the
quark couplings only, 
$A_{\rm LR,FB}^q = \frac{3}{4} {\cal P} \cAq$.
For this reason, a factor 15 improvement on $\cAb$ relative to the LEP/SLC
result is possible if polarised positrons are available. With polarised
electrons only the improvement is limited to a factor six due to the
polarisation error. To keep systematics under control also here the
improved b-tagging capabilities are essential.

\subsubsection{Observables at the WW-threshold}

The measurement of the W mass and width using a polarised threshold scan
at a linear collider has been investigated in \cite{GWW_WW}.
The study considered a dedicated scan with 100 fb$^{-1}$ of total integrated 
luminosity taken at several scan points near threshold with polarised
electron and positron beams.
It was demonstrated that the W mass could be measured with a precision of 5 MeV.
This error includes the statistical error and systematic errors arising
from event selection efficiency, background normalisation, luminosity normalisation
and absolute polarisation determination all of which can be determined, if needed,
from the data themselves.
Therefore with respect to these error sources, the mass determination is 
experimentally robust.
Point-to-point errors were assumed to be negligible; this seems a reasonable
assumption. Errors from uncertainties on the background spin model were explored; 
this topic deserves further study, but methods to alleviate and/or control such
an error as discussed in \cite{ref:orange} can be applied.

Nevertheless, in order to exploit this terrific potential, one will need to
control the absolute beam energy, the measurement of
the luminosity spectrum, and theoretical uncertainties of the
cross-section near 
threshold to sufficient precision.
Based on studies of solutions to these issues (for example as 
discussed in \cite{ref:orange}), it looks reasonable to be cautiously 
optimistic that such error sources can be controlled {\it by design}, leading to
an estimated overall error on the W mass of around 7 MeV
for an exposure of 100 fb$^{-1}$ (see also the discussion in~\cite{blueband}).

A similar scan would yield an error on the W-width of 3--4 MeV with
an additional systematic error
coming from the beam-spread which is 80 MeV for the TESLA
design. Hence if the beam-spread can be measured with an error of
order 1\%, the W-width could indeed be measured to around 4 MeV.

\subsection{Beam Energy Requirements}
\label{sec:ebeam}

To fully exploit the increased statistical precision available
for precision electroweak measurements at a future linear collider,
very precise determinations of the centre-of-mass collision
energy are required.
For the Z-pole running, this requirement is driven by the 
precision needed to convert the measured value of $\ALR$ into the 
theoretically useful quantity $\stl$.
An uncertainty in $\sqrt{s}$ of 3~MeV would contribute an additional
uncertainty to the determination of $\stl$ which would match the
total experimental precision expected using the Blondel scheme.
A determination of $\sqrt{s}$ at the level of ${\cal O}$(1~MeV) 
is therefore required to keep this uncertainty from dominating
the result.
At the W-pair threshold, a 10~MeV measurement of $\sqrt{s}$ would
lead to a 5~MeV uncertainty in $\MW$, again comparable to the
total expected experimental precision.

One strategy to achieve this level of precision at the Z-pole
is to calibrate the total cross-section observed at each scan
point against the known lineshape parameters measured at LEP.
This avoids the need to make an absolute measurement of $\sqrt{s}$
at the 1~MeV level, although having a device available to make
relative measurements to a similar precision is probably still
necessary from an operational standpoint.
This strategy also requires that the luminosity collected at
each scan point must be measured with a precision approaching 
that achieved at LEP1.
Clearly this technique will not work directly at the W-pair 
threshold, although the precise knowledge of $\MZ$ can be used
to help calibrate the beam energy measurement.

A spectrometer built with the same design and philosophy as
that employed at the SLC should be capable 
of achieving a $10^{-4}$ precision in $\sqrt{s}$, particularly 
if it can be cross-calibrated against the known value of $\MZ$ 
at the Z-resonance.
While this is just barely adequate for $\MW$, it is probably
inadequate for the precision required for $\ALR$, particularly
if the positrons are polarised.
One possibility for improving this precision is to use the
kinematics of M{\o}ller scattering off an internal gas jet
target.
This was studied for the beam energy measurement at LEP2,
and a precision approaching 2~MeV appeared 
reasonable\cite{Cecchi:1997ax}.
A detailed study of whether this approach is suitable for
a low repetition rate linear collider is still needed, however.
The kinematics of Compton scattering (potentially using
the available polarimeter infrastructure) is also a 
possibility which warrants further study.
A more detailed discussion of the requirements and prospects
for beam energy measurements can be found in Ref.~\cite{ebeam}.

\subsection{Physics opportunities at low energies}
\label{sec:physopp}

\subsubsection{Tests of the Standard Model}
\label{subsec:smtest}

Within the SM, the predictions for the electroweak precision observables 
are affected via loop corrections by contributions from 
the top quark mass, $m_t$, and the Higgs boson mass, $M_H$,
where the loop corrections to $\MW$ are usually contained in $\De
r$~\cite{deltar} (see Ref.~\cite{gigaz} for details.)
The effective leptonic weak mixing angle, $\sweff$, is defined through the
effective couplings $g_V^f$ and $g_A^f$ of the Z~boson to fermions at
the Z~resonance,
where the loop corrections enter through $g_{V,A}^f$ 
(see Ref.~\cite{gigaz} for details.)
The radiative corrections entering the predictions of $M_W$ and
$\sweff$ depend quadratically on $m_t$, while the leading
dependence on $M_H$ is only logarithmic. Comparing the theoretical
prediction of the electroweak precision observables with their
experimental value allows an indirect determination of 
the Higgs boson mass $M_H$. The uncertainty of
this indirect determination arises from the theoretical uncertainties
for the precision observables and from their experimental
errors~\cite{blueband}. 

The current theoretical uncertainties arising from uncertainties in
the input parameters are dominated by 
the uncertainty in the top quark mass, $\de m_t$, and the uncertainty
in the fine structure constant at the Z~boson mass scale,
$\de\De\al$. 
At a linear collider the top quark mass will be measured to better than 
200\,MeV. With this precision the uncertainty in the predictions due to the top
mass will be negligible. The status and prospects for $\De\al$ are reviewed
in detail in \cite{ref:jens}. Currently several methods to evaluate $\De \al$
are used all arriving at errors sufficient for the present data, but much too
large for the GigaZ precision. However, at least two of the methods 
discussed in \cite{ref:jens} have the potential to reduce the errors with
additional experimental measurements of the $\ee$ hadronic cross section
and theoretical progress on the quark masses far enough to compete with
the projected precision of GigaZ.
Also the uncertainty in the prediction of $\sweff$ due to the Z-mass
is similar to the experimental uncertainty with $10^9$ Z-decays. 

Assuming a future determination of $\De\al$ to a
precision of $\de\De\al = \pm 7 \times 10^{-5}$~\cite{deltaalphafut},
a top quark mass measurement down to $\de m_t = 130$~MeV and 
$\delta\alpha_s(M_Z) = 0.0010$ (from other GigaZ
observables~\cite{ref:marc}), the
future theory uncertainties including unknown higher-order corrections
are given by~\cite{blueband}
\BE
\de M_W({\rm theory}) = \pm 3.5~{\rm MeV}, \qquad
\de \sweff({\rm theory}) = \pm 3 \times 10^{-5} \qquad
({\rm future}) .
\label{futureunc}
\EE

The precisions for indirect determinations of $M_H$ are summarised in
Table~\ref{tab:indirectmh}, see Ref.~\cite{blueband} for details. It
becomes obvious that the inclusion of 
polarisation in combination with GigaZ drastically improves the
indirect $M_H$ determinations. By comparing the indirectly obtained
value to the then known experimental value allows for a stringent
test of the SM, becoming even more stringent by the use of
polarisation.

\begin{table}[htb!]
\renewcommand{\arraystretch}{1.5}
\BC
\begin{tabular}{|l||r|r||r|}

\cline{2-4} \multicolumn{1}{c||}{}
                                       & $M_W$ & $\sweff$ & all  \\  \hline
                                                                     \hline
now                                    &106 \% & 60 \%    & 58 \% \\ \hline
                                                                     \hline
LC (no low energy run, no pol.)        & 15 \% & 24 \%    & 14 \% \\ \hline
GigaZ (incl.\ polarisation)            & 12 \% &  8 \%    &  8 \% \\ \hline
\end{tabular}
\renewcommand{\arraystretch}{1}
\caption[] {\it\footnotesize
{Cumulative expected precisions for the indirect determination
of the Higgs boson mass, $\de M_H/M_H$, taking into account the
experimental errors and the theoretical uncertainties,
eq.~(\ref{futureunc})~\cite{blueband}. 
The last column shows the indirect Higgs boson mass determination from the
full set of precision observables.}   }
\label{tab:indirectmh}
\EC
\end{table}

\subsubsection{Tests of Supersymmetry}
\label{subsec:testsusy}

Supersymmetry (SUSY) is a prominent example of physics beyond the SM.
If low-energy SUSY is realized in nature, it
could well be discovered at the Tevatron or the LHC, and 
further explored at an $\ee$ LC.

Similarly to the case of the SM, the predictions for $M_W$ and $\sweff$
can also be employed for an indirect test
of the MSSM.  
In contrast to the Higgs boson mass in the SM, the lightest CP-even 
MSSM Higgs boson mass, $m_h$, is not a free parameter but can be
calculated from the other SUSY parameters. Comparing its prediction with
the experimentally measured value will allow to set further constraints
on the model. The precision observables $M_W$, $\sweff$ and $m_h$ are
particularly sensitive to the SUSY parameters of the scalar top and
bottom sector and of the Higgs sector. This could allow to
indirectly probe the masses of particles in supersymmetric theories
that might not be accessible directly neither at the LHC nor at the LC.

As a specific example of indirect informations obtainable from the
precision observables at GigaZ which could be complementary to direct
experimental measurements at RunII of the Tevatron, the LHC and the LC
the scalar top sector of the MSSM is considered here. If the lighter
scalar top quark is within the kinematical reach of the LC, the process
$e^+e^- \to \Stope \bar{\Stope}$ will allow to measure its mass,
$\mste$, and the mixing angle in the stop sector, $\cos \tst$,
with an accuracy of below the level of 1\%~\cite{lcstop}. 
These direct  
measurements can be combined with the indirect information on the mass
of the heavier scalar top quark, $\mstz$, from requiring consistency of the
MSSM (here the case of the unconstrained MSSM with real
parameters is considered) with a precise measurement of $M_W$, $\sweff$, 
and $m_h$.
The evaluation of $m_h$ has been performed with 
{\tt FeynHiggs}~\cite{feynhiggs} (based on Ref.~\cite{mhiggs2l}), while 
details about the evaluation and the theoretical errors of $M_W$ and
$\sweff$ can be found in Refs.~\cite{gigazlcws,gigazE3}. 

\begin{figure}[ht!]
\begin{center}
\includegraphics[height=8cm]{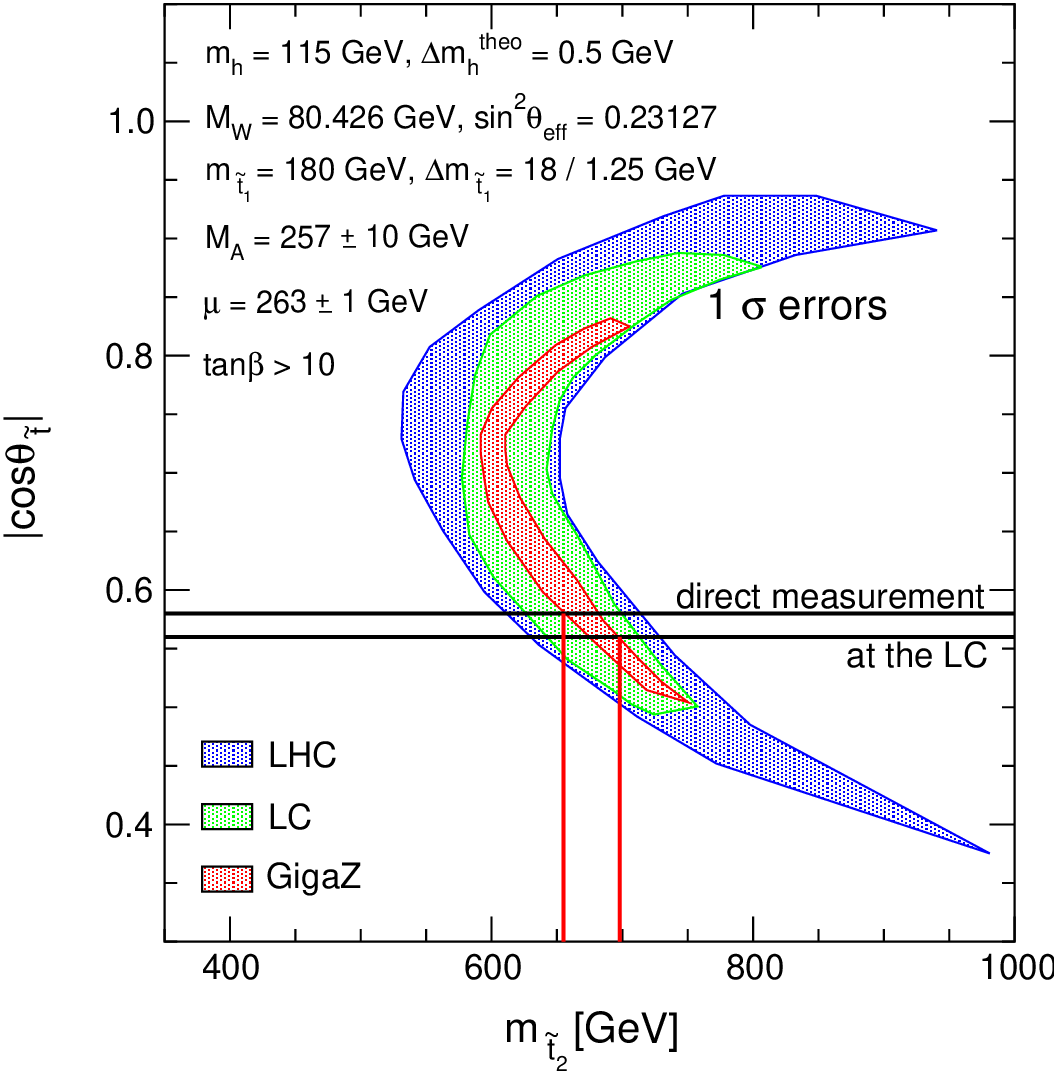}
\end{center}
\vspace{-1em}
\caption[]{
Indirect constraints on the MSSM parameter space in the 
$\mstz$--$|\cos \tst|$ plane from measurements of 
$m_h$, $M_W$, $\sweff$, $m_t$ and $\mste$ in view of the 
prospective accuracies for
these observables at an LC with and
without GigaZ option and at the LHC. The direct information on the
mixing angle from a measurement at the LC is indicated together with the
corresponding indirect determination of $\mstz$.
}
\label{fig:LCvGigaZ}
\end{figure}

Fig.~\ref{fig:LCvGigaZ}~\cite{gigazlcws} shows the allowed 
parameter space according to measurements of $m_h$, $M_W$ and $\sweff$
in the plane of the heavier stop mass, $\mstz$,
and $|\cos \tst|$ for the accuracies at an LC with
and without the GigaZ option and at the LHC. 
For $\mste$ the central value and experimental error
of $\mste = 180 \pm 1.25 \gev$ are taken for LC/GigaZ, while for the LHC an 
uncertainty of 10\% in $\mste$ is assumed, see Ref.~\cite{gigazlcws}
for details about the other SUSY parameters.
The central values for $M_W$ and $\sweff$ have been chosen in accordance
with a non-zero contribution to the precision observables from SUSY
loops. 

As one can see in Fig.~\ref{fig:LCvGigaZ}, the allowed parameter space in the
$\mstz$--$|\cos \tst|$ plane is significantly reduced
from the LHC to the LC, in particular in the GigaZ scenario. Using 
the direct information on $|\cos \tst|$ according to the analysis
performed in Ref.~\cite{lcstop}
allows an indirect determination of $\mstz$ with 
a precision of better than 5\% in the GigaZ case. By comparing this indirect 
prediction for $\mstz$ with direct experimental information on
the mass of this particle, the MSSM could be tested at its quantum level
in a sensitive and highly non-trivial way.

\end{document}